\documentclass[twocolumn,showpacs,preprintnumbers,amsmath,amssymb,prl,superscriptaddress]{revtex4}

\usepackage{graphicx}

\def\gz{\ifmmode{Z\hskip -4.8pt Z}
    \else{\hbox{$Z\hskip -4.8pt Z$}}\fi}

\newcommand{\be}{\begin{equation}}
\newcommand{\ee}{\end{equation}}
\newcommand{\bea}{\begin{eqnarray}}
\newcommand{\eea}{\end{eqnarray}}

\begin{document}

\title{Spin-charge separation in Aharonov-Bohm rings of interacting
electrons}

\author{K. Hallberg}
\affiliation{Centro At\'omico Bariloche and Instituto Balseiro,
Comisi\'on Nacional de Energ\'ia At\'omica, 8400 Bariloche, Argentina}

\author{A.~A. Aligia}
\affiliation{Centro At\'omico Bariloche and Instituto Balseiro,
Comisi\'on Nacional de Energ\'ia At\'omica, 8400 Bariloche, Argentina}

\author{A.~P. Kampf}
\affiliation{Institut f\"ur Physik, Theoretische Physik III,
Elektronische Korrelationen und Magnetismus, \\ Universit\"at Augsburg,
86135 Augsburg, Germany}

\author{B. Normand}
\affiliation{D\'epartement de Physique, Universit\'e de Fribourg,
CH-1700 Fribourg, Switzerland}


\begin{abstract}

We investigate the properties of strongly correlated electronic models 
on a flux-threaded ring connected to semi-infinite free-electron leads.
The interference pattern of such an Aharonov-Bohm ring shows sharp 
dips at certain flux values, determined by the filling, which are a 
consequence of spin-charge separation in a nanoscopic system. 

\end{abstract}


\maketitle

In interacting electron systems in reduced spatial dimensions, correlation
effects are strongly enhanced and the conventional quasiparticle description
of Fermi liquids may become inapplicable. In particular, for one-dimensional
(1d) systems the low-energy excitations are entirely collective in nature
and the Luttinger-liquid (LL) concept provides the appropriate framework 
to characterize the electronic properties. A hallmark of the LL is 
the fractionalization of the electronic excitations into separate collective 
spin and charge modes, a phenomenon known as spin-charge separation (SCS) 
\cite{Haldane,Schulz}.

With the advent of new materials and artificial structures of quasi-1d 
electronic character in the last decade, a variety of experiments has 
been reported which seek evidence of SCS. Prominent examples of candidate
materials include the organic Bechgaard and Fabre salts \cite{Bourbonnais},
molybdenum bronzes and chalcogenides \cite{Voit1}, cuprate chain and ladder
compounds \cite{DagottoRice}, and also carbon nanotube systems \cite{Byczuk}. 
The non-Fermi-liquid normal-state properties of high temperature 
superconductors have also led to attempts to trace their origin to the 
possible realization of SCS in strongly correlated electron systems in 2d 
\cite{Andersonbook}. Different approaches for the identification of SCS have 
included the analysis of non-universal power-law $I$-$V$ characteristics 
\cite{Voit1}, the search for characteristic dispersive features by 
angle-resolved photoemission spectroscopy (ARPES) \cite{Voit2}, 
establishing a violation of the Wiedemann-Franz law \cite{Taillefer}, 
and analyzing spin and charge conductivities \cite{Voit2,Qimiao}. While 
the interpretation of experimental results has been considered 
ambiguous in some cases, a verification of SCS has been reported from 
ARPES data on SrCuO$_2$ \cite{Kim}.

Theoretical methods for detecting and visualizing SCS were 
proposed and demonstrated many years ago. Direct calculations of the 
real-time evolution of electronic wave packets in Hubbard rings revealed 
that the spin and charge densities dispersed with different velocities as 
an immediate consequence of SCS \cite{Karen}. Equally striking was the 
analysis of transmission through Aharonov-Bohm (AB) rings \cite{Jagla}. 
The motion of the electrons in the ring was described by a LL propagator, 
where different charge and spin velocities, respectively $v_c$ and $v_s$, 
are included explicitly. With this assumption the flux-dependence of the 
transmission is no longer periodic only in multiples of a flux quantum 
$\Phi_0 = hc/e$, but instead new structures appear at fractional flux 
values which are determined by the ratio $v_s/v_c$. In essence, these 
structures arise because transmission requires the separated spin and 
charge degrees of freedom of an injected electron to recombine at the 
drain lead after traveling through the ring in the presence of the AB flux.

In this letter we propose an experimental configuration which may serve 
as a clean and direct probe of SCS. We employ the AB-ring transmission,  
focusing primarily on the $t$-$J$ model as a prototypical interacting 
system relevant to artificially designed 1d nanostructures such as small 
rings of quantum dots. We find a clear reduction of the transmittance of 
such a device at magnetic fields corresponding to fractional values of 
the flux in units of the flux quantum. These are explained in terms of 
an analysis of the momentum quantum numbers of the spin and charge
excitations, and of the orbital-flux-induced phase shifts accumulated on
traversing the ring between the two contact leads. The flux-dependent
interference pattern in the transmission through an AB ring is thus shown 
to be a valid and feasible tool for the unambiguous detection of certain 
signatures of the SCS phenomenon. 

\begin{figure}[t!]
\smallskip
\centerline{\includegraphics[height=5.0cm]{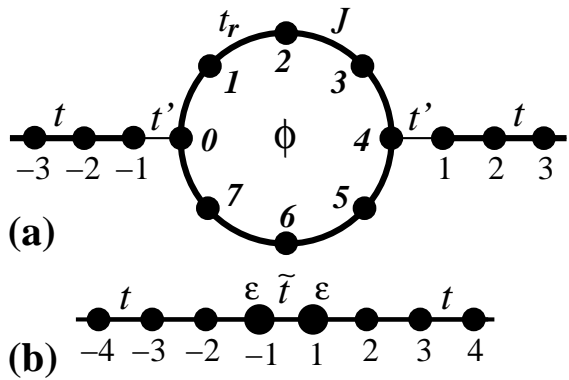}}
\vspace{3mm}
\caption{(a) Schematic representation of an interacting system on a ring 
connected by links $t^{\prime}$ to free-electron leads. The number of 
sites in the ring depicted is $L = 8$, and the transmittance is computed 
from the Green function connecting sites 0 and $L/2$. (b) Effective model 
of two impurities in a single conducting chain derived in the limit of 
weak $t^{\prime}$.}
\end{figure}

The system in Fig.~1(a) has the Hamiltonian 
\begin{equation}
H  = H_{{\rm leads}} + H_{{\rm link}} + H_{{\rm ring}},
\label{esh}
\end{equation}
where
\begin{equation}
H_{{\rm leads}} \! = \! -t \!\!\! \sum_{i=-\infty,\sigma}^{-1} \!\!\! a_{i-1, 
\sigma }^{\dag} a_{i,\sigma } \! -t \!\!\! \sum_{i=1,\sigma}^{\infty }\! a_{i,
\sigma }^{\dag} a_{i+1,\sigma } \! + \! {\rm H.c.} 
\label{ehl1}
\end{equation}
describes free electrons in the left and right leads, 
\begin{equation}
H_{{\rm link}} = -t^{\prime} \sum_{\sigma} (a_{-1,\sigma}^{\dag} c_{0,
\sigma} + a_{1,\sigma}^{\dag} c_{L/2,\sigma} + {\rm H.c.})  \label{ehl2}
\end{equation}
describes the exchange of quasiparticles between leads ($a_{i,\sigma }$)
and ring ($c_{l,\sigma }$), and 
\begin{eqnarray}
H_{{\rm ring}} & = &-e V_g \sum_{l,\sigma} c_{l,\sigma}^{\dag}c_{l,\sigma} 
- t_r \sum_{l,\sigma} (c_{l,\sigma}^{\dag} c_{l+1,\sigma} e^{-i\phi/L}
\nonumber \\ & & + {\rm H.c.}) + H_{{\rm int}}  \label{ehr}
\end{eqnarray}
describes the interacting electron system. The AB ring has length $L$, is 
threaded by flux $\phi$, where $\phi = 2\pi\Phi/\Phi_0$, and is subject to 
an applied gate voltage $V_g$.

Following Ref.~\cite{Jagla}, the transmission from the left to the right 
lead can be calculated to second order in $t^{\prime }$ from an effective 
low-energy Hamiltonian $H_{\rm eff}$ for the system with an additional 
particle of energy $\omega$ and wave vector $\pm k$ in the left or the 
right lead. $H_{\rm eff}$ is equivalent to the one-particle Hamiltonian 
for the chain represented in Fig.~1(b), with effective energy $\epsilon 
(\omega) = t^{\prime\,2} G_{0,0}^{{\rm R}}(\omega)$ for sites adjacent 
to the ring [$-1$ and 1 in Fig.~1(a)], and effective hopping $\tilde{t} 
(\omega) = t^{\prime \,2}G_{0,L/2}^{{\rm R}}(\omega )$ across the ring; 
$G_{i,j}^{{\rm R}}(\omega)$ denotes the Green function of the isolated ring.

At zero temperature, the transmittance and conductance of the system may 
then be computed using the effective impurity problem. This proceeds in 
the $T$-matrix formulation, which yields transmission amplitudes based 
on an intersite promotion matrix \cite{rsjc,rataf}. The transmittance 
$T(\omega)$ is given by \cite{Jagla} 
\begin{equation}
T(\omega,V_g,\phi) = \frac{4t^{2} \sin^{2} k |{\tilde{t}} (\omega)|^{2}}
{|[\omega - {\epsilon}(\omega) + te^{ik}]^{2} - |{\tilde{t}}^{2} 
(\omega)||^{2}}, \label{tra} 
\end{equation}
where $\omega = -2t \cos k$  is the tight-binding dispersion relation for 
the free electrons in the leads which are incident upon the impurities. 
These equations are exact for a non-interacting system, while with 
interactions on the ring they serve as an approximation in the tunneling 
limit $t^{\prime}/t \ll 1$ \cite{Jagla}. We comment that previous 
calculations of $T(\omega,V_g)$ in nanoscopic systems do not include 
interference effects which exist in a ring geometry. We emphasize that 
this analysis is applicable only for systems in which the ground state 
before and after particle or hole injection has no spin degeneracy. 
Thus Eq.~(5) does not include the Kondo effect, which arises in the 
spin-degenerate case, but in any event is destroyed even by small 
temperatures or applied magnetic fields.

To calculate $T(\omega,V_g)$ we employ numerical diagonalization of the 
isolated ring using a $t_r$-$J$ model where $H_{\rm int} = J \sum_{l} 
({\bf S}_{l} {\bf \cdot S}_{l+1} - 1/4)$ in Eq.~(\ref{ehr}), with 
${\bf S}_{l} = \sum_{\alpha\beta} c_{l\alpha}^{\dag} \sigma_{\alpha\beta} 
c_{l\beta}$ the spin at site $l$ and implicit projection of the $t_r$ term 
to single site occupancy. We consider particles incident on a ring of $L$ 
sites and $N+1$ electrons, obtaining the Green functions from the ground 
state of the isolated ring \cite{balseiro} and substituting these in 
Eq.~(\ref{tra}). We fix the energy $\omega = 0$ to represent half-filled 
leads and explore the dependence of the transmittance on the flux. 
$T(0,V_g)$ as a function of $V_g$ shows narrow peaks (with a width 
proportional to $(t^{\prime })^{2}$) at gate voltages which match the 
excitation energies of the system.

\begin{figure}[t!]
\hspace{-0.5cm}
\centerline{\includegraphics[width=7.8cm,angle=270]{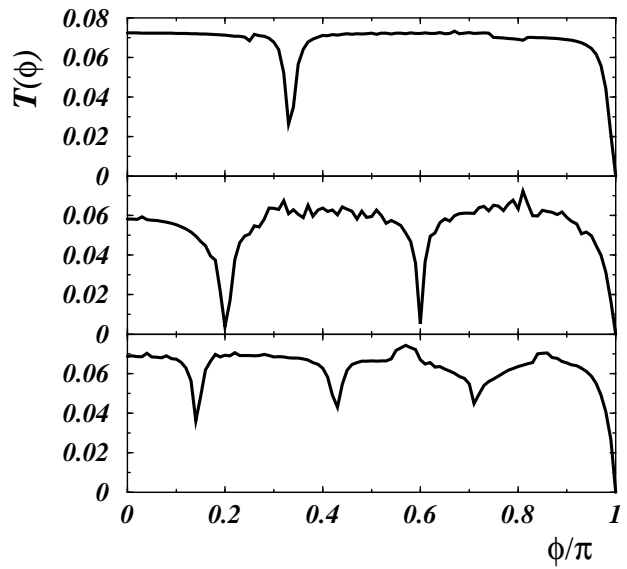}}
\caption{Transmittance as a function of flux for a $t_r$-$J$ model with $J = 
0.001t_r$, $t_r = t$, $t^{\prime} = 0.3t$, and $L = 8$ sites. The filling of 
the ring is (a) $N+1 = 4$, (b) $N+1 = 6$, and (c) $N+1 = 8$. The transmission 
occurs through intermediate states with $N = 3$, 5, and 7 particles, 
respectively, which lead to minima at flux values $\phi_d \approx \pi 
(1 - 2n_s/N)$ (see text).}
\label{fig:t-J1}
\end{figure}

The transmittance of the ring, obtained by integration over the 
excitations in a small energy window at the Fermi level \cite{Jagla}, 
is shown in Fig.~\ref{fig:t-J1} for different ring fillings. For 
injection of a hole in systems containing $N+1 = 4$, 6 or 8 particles 
[Fig.~\ref{fig:t-J1}(a)-(c)], the dynamical properties correspond to 
$N = 3$, 5, or 7 particles. The most striking result is the existence of 
dips at certain flux values, which constitute a clear signature of SCS.

A first obvious feature is that the transmittance vanishes at flux 
$\phi = \pi$. This is expected from negative interference of the 
components of the electron wave function traveling in the upper and 
lower halves of the ring. Formally, it is a consequence of the reflection 
symmetry of the device through the axis containing the leads. For $\phi = 
\pi $, the gauge transformation
\begin{equation}
f_{l,\sigma }^{\dag } = c_{l,\sigma }^{\dag } e^{il\phi /L}, \text{ } 
g_{m,\sigma}^{\dag } = a_{m,\sigma }^{\dag }e^{i\phi /2},m>0  \label{gauge}
\end{equation}
leads to a Hamiltonian with all hoppings real and of the same sign, except
between ring sites 0 and $L-1$, where it has the opposite sign. The 
transformed Hamiltonian is clearly invariant under simultaneous reflection 
and sign change of the phase of $c_{0,\sigma}^{\dag}$ ($c_{l,\sigma}^{\dag}
\leftrightarrow c_{L-l,\sigma }^{\dag }$ for $l > 0$, $c_{0,\sigma}^{\dag}
\leftrightarrow - c_{0,\sigma}^{\dag}$ for $l = 0$), whereas the Green 
function $G_{0,L/2}^{{\rm R}}(\omega)$ changes sign. The latter must be 
invariant with respect to operators of the symmetry group of $H$, and 
therefore vanishes along with $T(\omega )$ (\ref{tra}).

While the presence of the dip at $\phi = \pi$ is quite general, the origin
of the other dips in the transmittance resides in the strongly correlated
nature of the problem. We find numerically that the peaks are better 
defined for small interaction strengths ($J/t_r < 1/L$), as discussed  
below, and therefore begin our explanation of the presence of the 
additional dips by assuming $J = 0$. In this limit the model is equivalent 
to the infinite-$U$ Hubbard model, and complete SCS takes place on all 
energy scales \cite{ogata,rlw}.

Following the method of Ref.~\cite{caspers} for the ring with arbitrary flux, 
we construct spin wave functions which transform under the irreducible 
representations of the group $C_{N}$ of cyclic permutations of the $N$ 
spins of the $L$-site system. Each of these representations is labeled by 
a wave vector $k_{s} = 2 \pi n_{s}/N$, where the integer $n_{s}$ characterizes 
the spin wave function. For $J = 0$ each element of $C_{N}$ commutes with 
$H_{\rm ring}$. In each subspace of states whose spin wave function is 
characterized by the quantum number $k_{s}$, the problem may be mapped to 
that of a non-interacting, spinless system with effective flux \cite{caspers}
\begin{equation}
\phi_{\rm eff} = \phi + k_{s} = \phi + 2 \pi n_{s} / N .
\label{fief}
\end{equation}
The total energy of any state of the ring becomes
\begin{equation}
E = -2 t_{r} \sum_{l=1}^{N} \cos (k_{l} + \frac{\phi_{\rm eff}}{L}),\;\; 
k_{l} = \frac{2 \pi}{L} n_{l},  \label{ene}
\end{equation}
where $n_{l}$ and $N$ are charge quantum numbers. Thus the dynamical charge 
properties are described completely by a spinless model. The SCS phenomenon 
enters in that the spin wave function modifies the effective flux seen by 
the charges. Hole injection in an ($N$+1)-particle system (or particle 
injection for an ($N$-1)-particle system) yields an intermediate $N$-particle 
state with a certain weight for each spin wave number $k_{s}$, before the 
hole (particle) is ejected through the other lead. Because the charge 
dynamics in the ring are determined by the flux $\phi_{\rm eff}(k_{s})$, 
states with spin wave numbers $k_{s}$ do not contribute to the transport 
when $\phi_{\rm eff}(k_{s}) = \pi$ because they interfere destructively 
(above). From Eq.~(\ref{fief}) one therefore expects dips in the 
transmittance when $\phi = \phi_{d}$ with 
\begin{equation}
\phi_{d} = \pi (1 - 2 n_{s} / N)\, .  
\label{fid}
\end{equation}

\begin{figure}[t!]
\hspace{-0.5cm}
\centerline{\includegraphics[width=7.5cm,angle=270]{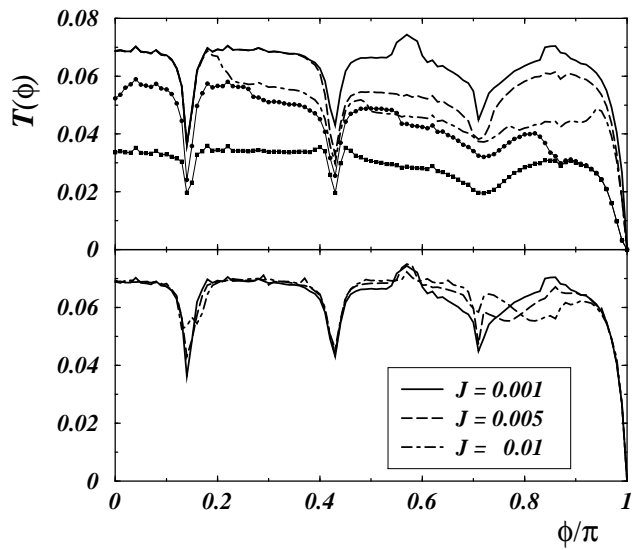}}
\caption{(a) Transmittance as a function of flux for a system of $L = 8$ 
with $J = 0.001 t_r$, $N+1 = 8$ particles and different window sizes ranging 
from $0.6t$ to $0.15t$ (top to bottom) below the Fermi energy. (b) As (a) 
for a selection of values of $J$.}
\label{fig:t-J2}
\end{figure}

In Fig.~2 we have considered the transmission contributions of excitations 
in a finite energy window, which accounts for possible (gate and bias) 
voltage fluctuations and temperature effects unavoidably present in an 
experimental system. For given $\phi_{d}(n_{s})$, the depth of the dip 
in the integrated transmittance depends on the number of energy levels 
which fall inside the window. If for given $V_g$ the window includes the 
destructively interfering states, then $T(\phi)$ exhibits dips at flux 
$\phi_d$. The dependence of the dip structures on the size of the window 
is shown in Fig.~3(a) for a system with $N+1 = 8$ particles. Although the 
integrated transmission decreases with window width, the principal features 
remain present, indicating the origin of the dips in destructive 
interference of levels very close to the Fermi energy. If the window size 
is too large (not shown), the dip depth diminishes due to the presence of 
additional, non-interfering peaks further from the Fermi level. 

Our numerical results indicate that the above reasoning remains valid 
for finite $J$, where SCS is incomplete. For $J = 0$, the lowest-energy 
states with consecutive charge quantum numbers \cite{caspers} include 
in general all possible $k_{s}$. We find a schematic correspondence of 
these states to effective ``spin'' subbands separated by a charge energy 
scale on the order of the finite-size gap of the ring, $t_r/L$. SCS is 
then expected to remain a valid concept for the spin and charge 
excitations of a small system for $J/t_r < 1/L$, a result confirmed 
by the progressive loss of well-defined dip structures in Fig.~3(b). The 
breakdown of SCS may then be described as an intrinsic phenomenon related 
to the mixing of different charge subbands. This mixing is strongest at 
higher values of the flux [Fig.~3(b)]. For small $J$ the position of the 
dips is shifted slightly, but remains close to the flux values given by 
Eq.~(\ref{fid}) for all fillings shown in Fig.~2. 

We summarize briefly the connection between the destructive interference 
of certain excited states, which is responsible for the transmittance 
dips, and the concept of separate, effective spin and charge velocities 
$v_s$ and $v_c$ in the ring, as observed by considering the evolution of 
electronic wave packets \cite{Karen}. In the limit $U \rightarrow \infty$, 
$v_s$ and $v_c$ can be obtained from the ratio of the change in total 
energy and momentum when either $n_s$ or one of the $n_l$ is increased 
by one, using either Eqs.~(7) and (8) or the Bethe-Ansatz equations of 
Ref.~\cite{ogata}. For the $N$-particle intermediate states relevant to 
our analysis, the effective spinless model has flux $\pi$ and odd $N$, 
circumstances under which for spin quantum number $n_s$ and flux 
$\phi_d (n_s)$ one obtains $v_s / v_c = 1/N$. This expression is then 
consistent with a qualitative understanding of the dips in Figs.~2(a)-(c) 
in terms of the charge and spin components of the injected particle 
performing different numbers of turns around the ring before reuniting 
at the drain lead. We stress, however, that in small $t_r$-$J$ and Hubbard 
rings a distribution in effective velocities for the states involved in 
transmission processes is unavoidable at arbitrary flux values. 

Transmittance dips should also be observable in an Aharonov-Casher (AC) 
geometry, where the ring is pierced by a charged wire \cite{aha,balatsky}, 
by extending the treatment to the case in which different fluxes act on 
particles with up- and down-spins \cite{afa}. Specifically, the mapping 
of charge degrees of freedom to a spinless model for $J = 0$ remains 
applicable with the substitution $\Phi \rightarrow \Phi + \sigma_z 
\Phi_{\rm AC}$ in Eq.~(\ref{fief}), where $\sigma_z$ is the spin 
projection on the axis of the ring and $\Phi_{\rm AC}$ is proportional 
to the radial electric field \cite{balatsky}. An experimental realization 
of the AB system requires the design of artificial structures, such as 
rings of quantum dots, on the sub-$\mu$m scale; accessible laboratory 
fields will not permit AB experiments on molecular rings. The wide 
variety of quantum-dot assemblies synthesized in recent years \cite{rkc} 
suggests that such structures are well within the compass of current 
nanofabrication technology \cite{rka}. Similar devices with a charged 
central gate could be used for measurements of the AC effect, which may 
also be feasible on a more molecular scale with multi-wall nanotubes, 
charged nanotubes piercing large molecular rings, or closed-loop 
nanotube assemblies.

In conclusion, the transmittance through AB rings of interacting 
electrons provides a straightforward technique for the detection of SCS. 
The existence of transmission dips arising from non-trivial destructive 
interference effects at fractional values of the flux quantum $\Phi_0$ 
is a robust signature of SCS. While the depths and widths of the dips 
vary with the system under consideration, their positions depend only 
on ring filling and weakly on interaction strengths. The experimental 
capability to construct systems exhibiting nanoscopic SCS exists already. 

\smallskip 

We are grateful to G. Japaridze and M. Sekania for helpful discussions. 
K.H. and A.A.A. are fellows of CONICET, and thank the Fundaci\'{o}n 
Antorchas, Project 14116-168, and PICT 03-12742 of ANPCyT for support. 
A.P.K. acknowledges the support of the Deutsche Forschungsgemeinschaft 
through SFB 484 and B.N. the support of the Swiss National Science 
Foundation.

\end{document}